\documentclass[prl,reprint,twocolumn,showpacs,superscriptaddress]{revtex4-2}
\usepackage{graphicx}
\usepackage{dcolumn}
\usepackage{bm}
\usepackage{graphicx}
\usepackage{epsfig}
\usepackage{epsf}
\usepackage{amssymb}
\usepackage{amsmath,empheq,cases}
\usepackage{amsthm}
\usepackage{mathtools}
\usepackage{multirow}
\usepackage[colorlinks=true,linkcolor=blue,citecolor=blue,urlcolor=blue,pdfauthor={ },pdftitle={ },pdfsubject={ },pdfkeywords={ }]{hyperref}
\usepackage{pgfplots}
\usepackage{lipsum}

\usepackage{times}

\usepackage{color}
\definecolor{dred}{rgb}{.8,0.2,.2}
\definecolor{ddred}{rgb}{.8,0.5,.5}
\definecolor{dblue}{rgb}{.2,0.2,.8}
\definecolor{dgreen}{rgb}{.2,0.5,.2}

\usepackage{tikz}
\definecolor{c2}{RGB}{153,51,51}

\newcommand{\ket}[1]{|{#1}\rangle}

\usepackage[colorinlistoftodos,bordercolor=red!50,backgroundcolor=red!10,linecolor=red!50,textsize=tiny]{todonotes}
\usepackage{changes}

\newcommand\redsout{\bgroup\markoverwith{\textcolor{red}{\rule[0.5ex]{2pt}{1pt}}}\ULon}

\begin{document}
	\title{Self-Consistent Determination of Single-Impurity Anderson Model Using Hybrid Quantum-Classical Approach on a Spin Quantum Simulator}
	
	\author{Xinfang Nie}
	\thanks{These authors contributed equally to this work.}
	\affiliation{Shenzhen Institute for Quantum Science and Engineering and Department of Physics, Southern University of Science and Technology, Shenzhen 518055, China}	
\affiliation{Quantum Science Center of Guangdong-Hong Kong-Macao Greater Bay Area, Shenzhen 518045, China}
	
	\author{Xuanran Zhu}
	\thanks{These authors contributed equally to this work.}
	\affiliation{Shenzhen Institute for Quantum Science and Engineering and Department of Physics, Southern University of Science and Technology, Shenzhen 518055, China}
	\affiliation{Department of Physics, The Hong Kong University of Science and Technology, Clear Water Bay, Kowloon, Hong Kong, China}

	\author{Yu-ang Fan}
	\affiliation{Shenzhen Institute for Quantum Science and Engineering and Department of Physics, Southern University of Science and Technology, Shenzhen 518055, China}
	
\author{Xinyue Long}
	\affiliation{Shenzhen Institute for Quantum Science and Engineering and Department of Physics, Southern University of Science and Technology, Shenzhen 518055, China}
\affiliation{Quantum Science Center of Guangdong-Hong Kong-Macao Greater Bay Area, Shenzhen 518045, China}

	\author{Hongfeng Liu}
	\affiliation{Shenzhen Institute for Quantum Science and Engineering and Department of Physics, Southern University of Science and Technology, Shenzhen 518055, China}

	\author{Keyi Huang}
	\affiliation{Shenzhen Institute for Quantum Science and Engineering and Department of Physics, Southern
University of Science and Technology, Shenzhen 518055, China}	
	
\author{Cheng Xi}
	\affiliation{Shenzhen Institute for Quantum Science and Engineering and Department of Physics, Southern University of Science and Technology, Shenzhen 518055, China}

\author{Liangyu Che}
	\affiliation{Shenzhen Institute for Quantum Science and Engineering and Department of Physics, Southern University of Science and Technology, Shenzhen 518055, China}

\author{Yuxuan Zheng}
	\affiliation{Shenzhen Institute for Quantum Science and Engineering and Department of Physics, Southern University of Science and Technology, Shenzhen 518055, China}

\author{Yufang Feng}
	\affiliation{Shenzhen Institute for Quantum Science and Engineering and Department of Physics, Southern University of Science and Technology, Shenzhen 518055, China}

\author{Xiaodong Yang}
	\affiliation{Shenzhen Institute for Quantum Science and Engineering and Department of Physics, Southern University of Science and Technology, Shenzhen 518055, China}

	\author{Dawei Lu}
	\email{ludw@sustech.edu.cn}
	\affiliation{Shenzhen Institute for Quantum Science and Engineering and Department of Physics, Southern University of Science and Technology, Shenzhen 518055, China}
\affiliation{International Quantum Academy, Shenzhen 518055, China}
\affiliation{Quantum Science Center of Guangdong-Hong Kong-Macao Greater Bay Area, Shenzhen 518045, China}

\date{\today}
\begin{abstract}
The accurate determination of the electronic structure of strongly correlated materials using first principle
methods is of paramount importance in condensed matter physics, computational chemistry,
and material science. However, due to the exponential scaling of computational resources, incorporating
such materials into classical computation frameworks becomes prohibitively expensive. In 2016,
Bauer et al. proposed a hybrid quantum-classical approach to correlated materials [\href{https://journals.aps.org/prx/abstract/10.1103/PhysRevX.6.031045}{Phys. Rev. X \textbf{6},
031045 (2016)}] that can efficiently tackle the electronic structure of complex correlated materials. Here,
we experimentally demonstrate that approach to tackle the computational challenges associated with
strongly correlated materials. By seamlessly integrating quantum computation into classical computers,
we address the most computationally demanding aspect of the calculation, namely the computation
of the Green's function, using a spin quantum processor. Furthermore, we realize a self-consistent
determination of the single impurity Anderson model through a feedback loop between quantum and
classical computations. A quantum phase transition in the Hubbard model from the metallic phase to the Mott insulator is
observed as the strength of electron correlation increases. As the number of qubits with high control
fidelity continues to grow, our experimental findings pave the way for solving even more complex
models, such as strongly correlated crystalline materials or intricate molecules.
\end{abstract}

\maketitle

\emph{Introduction.}---Density-functional theory (DFT) stands as a pivotal cornerstone in contemporary materials simulation, spanning applications from molecules to solids~\cite{PhysRev.136.B864, RevModPhys.71.1253,RevModPhys.87.897,RevModPhys.78.865}. DFT efficiently mitigates the computational resources required by recasting the solution to the ground state of the Schr{\"o}dinger equation as an energy minimization problem dependent on the charge density. Nevertheless, its reliability diminishes when applied to strongly correlated systems, where the assumption of independent electrons no longer holds~\cite{RevModPhys.70.1039,kovaleva2008versatility}. Dynamical mean field theory (DMFT)~\cite{RevModPhys.78.865}, a non-perturbative approach, sheds light on the electronic structure of strongly correlated systems through a self-consistent loop~\cite{PhysRevB.45.6479,RevModPhys.68.13}. By mapping the many-body lattice problem to a many-body local problem, specifically the impurity model, computational complexity is substantially reduced. The integration of DFT and DMFT has emerged as a potent first-principle approach (DFT+DMFT) for tackling strongly correlated materials \cite{anisimov1997first,PhysRevB.57.6884,PhysRevB.80.235104,PhysRevLett.115.256402,PhysRevB.73.155112,
PhysRevB.85.235136,PhysRevB.102.245104,PhysRevB.94.195146,PhysRevLett.112.146401,PhysRevLett.120.187203}. In this framework, a computationally efficient DFT calculation is employed to establish a set of orbitals and determine the electronic structure for the majority of the orbitals, while a more computationally intensive DMFT is embedded to solve a simplified impurity model involving a significantly smaller set of correlated orbitals.

\begin{figure}
  \centering
  \includegraphics[width=0.45\textwidth]{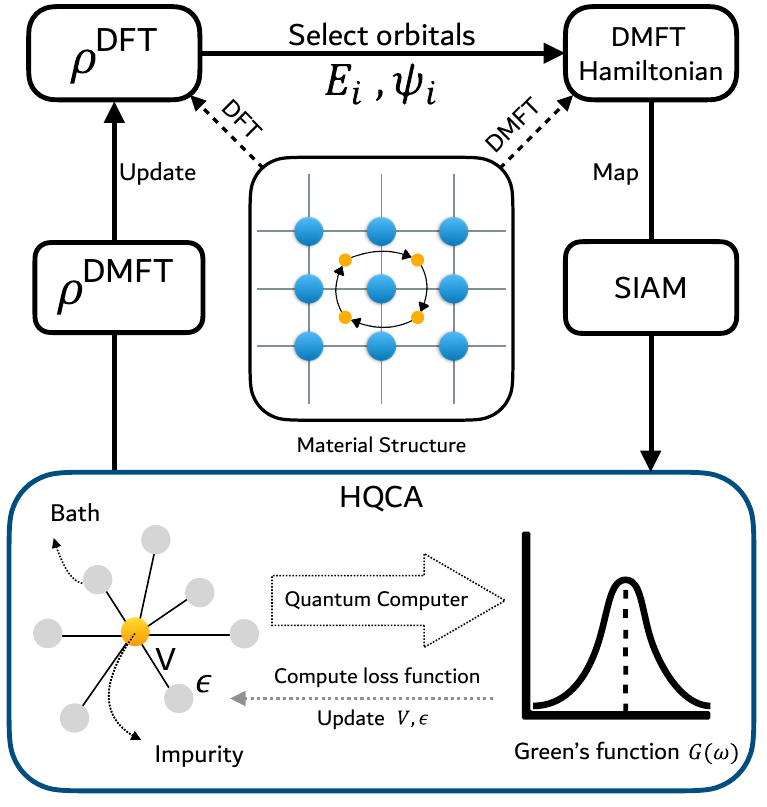}
  \caption{Flowchart of the DFT+DMFT process. This self-consistent loop follows an iterative procedure with the quantum computer serves as the impurity solver, as illustrated in the HQCA box. }\label{Fig1}
\end{figure}

However, calculating Green's function in the impurity model becomes notably challenging as the number of correlated orbitals increases. The essential tasks involved are determining the ground state and simulating the quantum evolution of the impurity model. These steps become increasingly intractable with classical methods when dealing with a larger number of correlated orbitals, owing to the heightened computational complexity~\cite{schuch2009computational,bravyi2017complexity,supple}.
Fortunately, quantum computers offer a potential solution to this problem \cite{aspuru2005science,cheng2023noisy}. In 2016, a hybrid quantum-classical approach (HQCA) was developed~\cite{PhysRevX.6.031045}, seamlessly merging classical and quantum algorithms into the DFT+DMFT framework as depicted
in Fig.~\ref{Fig1}. Quantum computers are leveraged to tackle the impurity problem more efficiently and accurately, while the computationally inexpensive components are executed on classical computers. HQCA~\cite{zhu2022calculating} is notably well-suited for noisy intermediate-scale quantum devices. Relevant algorithms have already found application in determining molecular ground states \cite{PhysRevX.8.031022}, simulating quantum dynamics \cite{PhysRevX.7.021050,PhysRevLett.125.010501,google2020hartree}, quantum machine learning~\cite{PhysRevLett.126.110502}, and optimizing quantum controls \cite{PhysRevLett.118.150503,PhysRevLett.123.130501,lin2023online}. These developments appear to be the most promising avenue for harnessing the benefits of quantum computing.

In this work, we present a practical demonstration of the HQCA for addressing the computational complexities inherent in strongly correlated materials. Our experiment achieves a self-consistent determination of the single-impurity Anderson model (SIAM), incorporating one bath site, through a feedback loop that combines quantum and classical computations. The quantum component is executed on a five-qubit nuclear magnetic resonance (NMR) quantum processor.
Experimentally, we measure the Green's function of the model across a spectrum of electron interactions. We iteratively determine the hopping matrix connecting the impurity to the bath, enhancing our understanding of the system's dynamics. Furthermore, our analysis reveals the quantum phase transition from a metallic state to a Mott insulator in the famous Hubbard model on a Bethe lattice, a phenomenon discerned by examining the spectral function.


\emph{Model.}---DMFT ingeniously transforms the complexity of a many-body lattice problem into a more tractable many-body local problem, commonly referred to as the impurity model. The simplest SIAM with just one impurity site and one bath site \cite{PhysRevA.92.062318,PhysRevB.64.165114,kreula2016EPJQuanTech} allows us to express the Hamiltonian as
\begin{equation}\label{Eq1}
\begin{aligned}
\mathcal{H}_\text{SIAM} = &U\hat{n}_{1\downarrow} \hat{n}_{1\uparrow}- \mu\sum_{\sigma} \hat{n}_{1\sigma} + \epsilon \sum_{\sigma} \hat{n}_{2\sigma}\\
 &+ V \sum_{\sigma} (\hat{c}_{1\sigma}^\dagger \hat{c}_{2\sigma}+h.c.),
\end{aligned}
\end{equation}
where the subscript ``1'' denotes the impurity site and ``2'' denotes the bath site. The symbol \(\sigma = \{\downarrow, \uparrow\}\) denotes the spin states associated with these sites.  The term \(U\geq 0\) signifies the interaction strength between electrons at the impurity site, while \(\mu\) represents the chemical potential. Furthermore,
\(\epsilon\) defines the bath energy, and \(V\) is the bath coupling that allows hopping between the sites. The operators \(\hat{c}_{1\sigma}^\dagger\) and \(\hat{c}_{2\sigma}^\dagger\) create fermions at the respective sites, with $h.c.$ denoting the Hermitian conjugate.

To experimentally simulate the SIAM within a spin system, we consider mapping the half-filled problems to the SIAM, i.e., \(\mu=U/2\). We fix \(\epsilon=0\) in Eq.~(\ref{Eq1}), and subsequently employ the Jordan-Wigner transformation to map the SIAM onto the Pauli basis~\cite{kreula2016EPJQuanTech,PhysRevLett.63.322}. This transformation yields the corresponding Hamiltonian
\begin{equation}\label{Eq2}
\mathcal{H}=\frac{U}{4} \sigma_z^1 \sigma_z^3+\frac{V}{2}\left(\sigma_x^1 \sigma_x^2+\sigma_y^1 \sigma_y^2+\sigma_x^3 \sigma_x^4+\sigma_y^3 \sigma_y^4\right),
\end{equation}
where \(\sigma_{x,y,z}^l\) is the corresponding Pauli operator for qubit \(l\) (\(l = 1, 2, 3, 4\)).
The interaction strength \(U\) is determined by the properties of the underlying material, while the bath coupling \(V\) must be established through a self-consistent loop within the DMFT framework.

The self-consistent loop, as illustrated in Fig.~\ref{Fig1}, follows an iterative procedure:
(i) An initial guess of the bath coupling \(V\) is loaded.
(ii) The ground state of the SIAM in Eq.~(\ref{Eq2}) is solved. After that, the Green's function of the impurity model, denoted as \(G\), is computed.
(iii) The approximate noninteracting impurity Green's function \(\widetilde{G}_0\) is derived.
(iv) The cost function,
\begin{equation} \label{cost}
f=\left|\widetilde{G}_0^{-1}-G_0^{-1}\right|^2
\end{equation}
is calculated, where \(G_0\) is the exact noninteracting Green's function determined by \(V\). (v) \(V\) is updated, and the iterative process continues with steps (ii) to (iv) until the value of \(f\) converges.

The significant computational expense is associated with the impurity solver, specifically, the computational step denoted as (ii) above. In the HQCA, the quantum computer functions as the impurity solver, while the remaining computational tasks are executed on a classical computer. Therefore, efficient measurement of the impurity's Green's function is the central component of the HQCA, which will be our primary focus in the subsequent discussion~\cite{supple}.

The particle (p) and hole (h) Green's functions of the SIAM in real-time can be expressed as~\cite{supple}
\begin{equation}\label{Eq3}
G^{\mathrm{p,h}}(t)=[O_1(t)\pm O_2(t)]/2,
\end{equation}
with $O_1(t)$ and $O_2(t)$ being experimental observables:
\begin{equation}\label{Eq4}
\begin{aligned}
O_1(t) & =\langle\Psi\vert e^{i \mathcal{H} t} \sigma_x^1 e^{-i \mathcal{H} t} \sigma_x^1\vert \Psi\rangle, \\
O_2(t) & =\langle\Psi\vert e^{i \mathcal{H} t} \sigma_x^1 e^{-i \mathcal{H} t}\left(-\sigma_y^1\right)\vert \Psi \rangle.
\end{aligned}
\end{equation}
Here, $\vert\Psi\rangle$ is the ground state of the Hamiltonian given by Eq.~(\ref{Eq2}). The Green's function in the frequency domain can be computed from Eq.~(\ref{Eq3}) as
\begin{equation}\label{Eq5}
G(\omega)=-i \int_\varepsilon^{\infty} e^{i(\omega+i \eta) t}\left[G^\text{p}(t)+\bar{G}^\text{h}(t)\right]dt ,
\end{equation}
where $\varepsilon$ and $\eta$ are chosen for numerical stability, and $\bar{~}$ represents the complex conjugate. Hence, we can compute the cost function $f$ in Eq. (\ref{cost}), which facilitates the self-consistent loop in determining the bath coupling $V$.

\begin{figure*}
  \centering
  \includegraphics[width=0.95\textwidth]{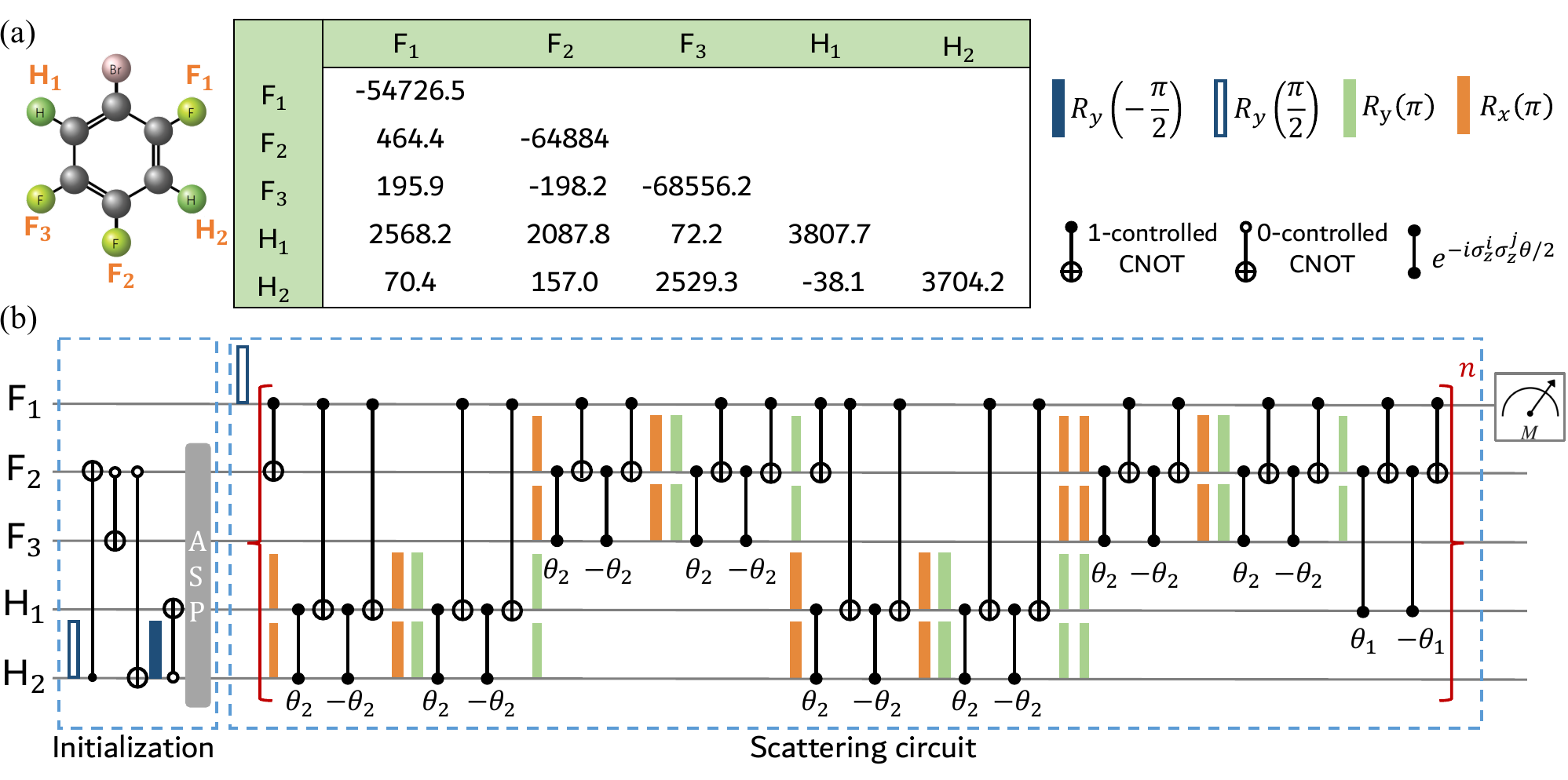}
  \caption{(a)  Molecular structure and parameters for the five-qubit NMR sample, composed of three $^{19}$F and two $^1$H spins identified as qubits. The table presents chemical shifts (diagonal elements) and coupling strengths (off-diagonal elements), both in Hertz. (b) Experimental quantum circuit for measuring the correlation $O_1(t)$, with $\theta_1=Ut/4n$ and $\theta_2=Vt/2n$. Here, the first-order Trotter approximation with $n=1$ is employed.
 }\label{Fig2}
\end{figure*}

\emph{Experimental setting.}---
To directly measure $O_1(t)$ and $O_2(t)$ in the experiment, we employ the quantum scattering circuit~\cite{PhysRevLett.81.5672,PhysRevA.105.L030402}. By introducing one ancillary qubit, this approach avoids full quantum state tomography. Hence, the experiment involves five qubits, where four are used to simulate the SIAM and one ancilla to perform the measurement. The five-qubit quantum processor comprises two $^1$H nuclear spins and three $^{19}$F nuclear spins within 1-bromo-2,4,5-trifluorobenzene \cite{SHANKAR201410,PhysRevA.90.012306}, dissolved in liquid crystal solvent  MBBA.  The spins F$_1$, F$_2$, F$_3$, H$_1$, and H$_2$, are designated as qubits 1 through 5, respectively; see Fig.~\ref{Fig2}(a).  All the experiments are carried out on a Bruker 600 MHz spectrometer at 305 K.

The quantum scattering circuit is depicted in Fig.~\ref{Fig2}(b).
The first spin serves as the ancillary qubit. The second and third spins represent the $\ket{\downarrow}$ state for both fermionic sites, while the fourth and fifth spins correspond to the $\ket{\uparrow}$ state.
The quantum simulation proceeds in the following three steps: (i) Initialize the quantum simulator to the state $\vert0\rangle\otimes\vert\Psi\rangle$ through adiabatic evolution.
(ii) The five-qubit system undergoes the controlled evolution of the SIAM, represented by $\mathcal{U}_\text{C}=\vert 0\rangle\langle0\vert\otimes\bm{I}+\vert1\rangle\langle1\vert\otimes \mathcal{U}_{1,2}$. Here, $\bm{I}$ is the $16\times 16$ identity matrix. The choice of $\mathcal{U}_{1,2}$ depends  on the measurement target, either $O_1(t)$ or $O_2(t)$. Specifically, for $O_1(t)$, $\mathcal{U}_1=e^{i\mathcal{H}t}\sigma_x^1e^{-i\mathcal{H}t}\sigma_x^1$ is used, and for $O_2(t)$, $\mathcal{U}_2=e^{i\mathcal{H}t}\sigma_x^1e^{-i\mathcal{H}t}(-\sigma_y^1)$ is used; see Eq. (\ref{Eq4}). (iii) Obtain $O_{1,2}(t)$ by directly reading out $\langle\sigma_x^1 \rangle$ of the ancillary qubit.

\begin{figure*}[htb]
\centering
\includegraphics[width=0.95\textwidth]{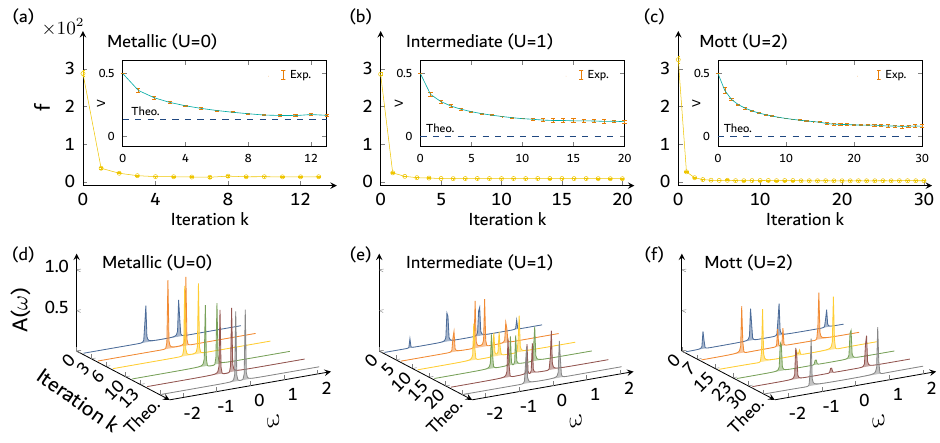}
\caption{Experimental results of the iterative process for three different interaction strengths: $U=0$, $1$ and $2$.
(a-c) Changes in the cost function $f$, and bath coupling $V$ with increasing iteration $k$. The initial values of the bath couplings are uniformly set at $V_0 = 0.5$.
For each iteration $k$, the experiment is repeated four times to derive the statistical error bars.
The dashed lines represent the theoretical values of $V$ obtained from the classical simulation.
(d-f) Experimental spectral density function $A(\omega)$ as a function of iteration $k$ for the three cases, alongside the corresponding final theoretical spectral density functions represented by gray lines.
}\label{Fig3}
\end{figure*}

The crux of the experiment lies in preparing the five-qubit system in its initial state $\vert0\rangle\otimes\vert\Psi\rangle$. The system begins with $\vert 00000\rangle$, which we prepare using an
improved line-selective method~\cite{peng2001preparation}. Our new method designs the shape pulses analytically, eliminating the need for numerical optimization and precisely designed shaped pulses to remove the zeroth-order quantum coherences. For further details, see the Supplementary Material~\cite{supple}. To achieve adiabatic state preparation~\cite{PhysRevLett.104.030502,PhysRevLett.108.130501}, we design a passage connecting a ground state of a simple Hamiltonian to the target Hamiltonian's ground state. We select the initial state as $\vert\Psi_0\rangle=(|0101\rangle-|0110\rangle-|1001\rangle+|1010\rangle)/2$, the SIAM ground state with $U = 0$ and any positive $V$. $\vert\Psi_0\rangle$ can be effectively prepared from $\vert0000\rangle$ using line-selective operations and is independent of $V$ provided that $V$ is positive. By slowly increasing $U$  to satisfy the adiabatic condition, the SIAM ground state for any  $U$ and $V$~\cite{supple}. The optimization of adiabatic state preparation is achieved through a gradient-based optimization technique~\cite{khaneja2005optimal}.

In addition to ground-state initialization, a significant challenge in our experiment is simulating the controlled unitary operation $\mathcal{U}_\text{C}$. This operation can be decomposed into controlled single-qubit rotations and controlled coupling evolutions. For instance, considering the case $\mathcal{U}_\text{C}=\vert0\rangle\langle0\vert\otimes\bm{I}+\vert1\rangle\langle1\vert\otimes \mathcal{U}_1$, the gate can be decomposed as $\mathcal{U}_\text{C}=u_1^\dagger u_2u_1u_2$, where $u_1=\vert0\rangle\langle0\vert\otimes\bm{I}+\vert1\rangle\langle1\vert\otimes e^{-i\mathcal{H}t}$ and $u_2=\vert0\rangle\langle0\vert\otimes\bm{I}+\vert1\rangle\langle1\vert\otimes \sigma_x^1$.
The evolution $e^{-i\mathcal{H}t}$ can be realized by the Trotter-Suzuki formula~\cite{suzuki1985decomposition}, see the Supplementary Material for details~\cite{supple}. The quantum circuit of the experiment is depicted in Fig.~\ref{Fig2}(b).
Experimental control accuracy is further enhanced by engineering a shaped pulse via gradient-based optimization~\cite{khaneja2005optimal}. In experiment, the length of the shaped pulse for $\mathcal{U}_\text{C}$ is 40 ms with a simulated fidelity exceeding 0.995.

\emph{Self-consistent loop.}---The core of the HQCA to simulate the strongly correlated model lies in iteratively determining the self-consistent value of $V$ using the quantum processor. The experiment operates as follows: For a fixed value of the interaction strength $U$, we commence with an initial estimate of the bath coupling, denoted as $V_0$, where the subscript represents the iteration number. We prepare the corresponding initial state and then measure the real-time $O_{1,2}(t)$ functions defined in Eq.~(\ref{Eq4}) using the quantum scattering circuit. Subsequently, we numerically compute the cost function $f_0$ by combining Eqs. (\ref{cost}-\ref{Eq5}) from the experimental data on a classical computer.

The gradient of the cost function, denoted as $g=\nabla f$,
is determined using the finite-difference method by evaluating the change in $f$   resulting from a small step $\delta V = 0.05$~\cite{supple}.
In the $k$-th iteration, we adjust $V_k$ by $\delta V$, repeat the experimental circuit to measure $O_{1,2}(t)$, and then compute the new value of the cost function.This allows us to calculate $g$ and update $V_k$ to $V_{k+1}$ for the subsequent iteration. The optimization continues until $f$ converges to a minimum value and $V$ reaches an optimal and stable value. Concurrently, the spectral density corresponding to each $V_k$ is computed throughout the optimization process.

\emph{Results.}---We select the half-filled Hubbard model on a Bethe lattice as our test ground. The experimental results of the iteration process for three different interaction strengths $U=0, 1, 2$ are presented in Fig.~\ref{Fig3}. These results reveal a distinct convergence pattern in both the cost function $f$ [see Figs. \ref{Fig3}(a-c)] and the bath coupling $V$  [see the insets in Figs. \ref{Fig3}(a-c)] across all three cases, serving as a signature of the successful solution of the Hubbard model.
The initial bath couplings are uniformly set at $V_0 = 0.5$ for all three cases. After 13, 20, and 30 iterations of the HQCA optimization, the cost function $f$ gradually converged to relatively small values. At the end of the optimization, the final bath coupling strengths are determined to be $V = 0.171\pm0.005, 0.121\pm0.013$, and $0.082\pm0.011$ for the respective cases~\cite{supple}. The bath coupling $V$ decreases as the interaction strength $U$ increases, which is consistent with previous studies.

Furthermore, the spectral density function, defined as $A(\omega)=-\text{Im}G(\omega)/{\pi}$, is displayed in Figs. \ref{Fig3}(d-f) for the three interaction strengths. For $U = 0$, the two peaks in the spectral density gradually shift towards lower frequency range as iterations progress, ultimately stabilizing and suggesting the emergence of a Fermi liquid state (metallic phase). In contrast, for $U = 2$, the peaks move in the opposite direction, indicative of a transition towards a Mott insulator phase.
For $U = 1$, two peaks are initially observed, which, by the end of the iteration,evolve into four distinct peaks, hinting at an intermediate state between the Fermi liquid and Mott insulator.

\begin{figure}
\centering
\includegraphics[width=0.48\textwidth]{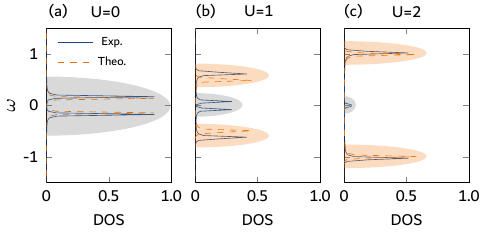}
\caption{DOS for three different interaction strengths, $U = 0$, $1$, and $2$. The solid and dashed peaks represent the experimental and theoretical results for the SIAM model, respectively.  The shaded regions illustrate the schematic energy bands of the metal and insulator when we use the SIAM with many bath sites.}\label{Fig4}
\end{figure}

Let us focus on the final density of states (DOS) for the three cases, as shown in Fig.~\ref{Fig4}.
For the case of $U = 0$, where electrons are completely independent, the experimental results exhibit two symmetrically distributed peaks near $\omega=0$, as shown in Fig.~\ref{Fig4}(a). For the case of $U = 1$, the spectrum exhibits a characteristic structure of the Hubbard bands, which originates from local atomic excitations and is broadened by the hopping of electrons away from the atom, and the quasiparticle peak near the Fermi level, as shown in Fig.~\ref{Fig4}(b). When the electron interactions are sufficiently strong ($U = 2$), it causes the quasiparticle peak to vanish, as the spectral weight of that low-frequency peak is transferred to the high-frequency Hubbard bands, as shown in Fig.~\ref{Fig4}(c). The presence of discrete peaks can be attributed to the limited number of bath sites in the experiment~\cite{PhysRevB.90.085102} while the ideal DOS according to the SIAM with many bath sites  is also illustrated by the shadowed areas for each case. In conclusion, the results provide a qualitative experimental observation of a Mott transition occurring from a Fermi liquid to a Mott insulator as $U$ increases from $U=0$ to $U=2$.

\emph{Conclusion.}---
In this work, we present a quantum impurity solver utilizing the HQCA to address strongly correlated models. In the realm of computational physics, DMFT is a widely adopted tool for deciphering the electronic structure of strongly correlated materials. However, within a classical framework, DMFT becomes computationally expensive, constraining the DFT+DMFT method to single impurity models with a limited number of correlated orbitals.

As a promising candidate for near-term practical applications in quantum computing, the HQCA offers a solution to overcome these limitations. Here, the quantum processor takes on the role of the impurity solver. In our endeavor, we take the first step by experimentally solving a strongly correlated model using an NMR quantum processor to ascertain the corresponding bath couplings of the SIAM.
Despite the constraint of a limited number of bath sites, our experiment provides preliminary evidence of a quantum phase transition from the metallic phase to a Mott insulator for the famous Hubbard model on a Bethe lattice. This work opens the door to harnessing the immense potential of quantum computing for tackling the complexities of strongly correlated materials in computational physics.

\begin{acknowledgments}
This work is supported by the National Key Research and Development Program of China (2019YFA0308100), the National Natural Science Foundation of China (12104213,12075110,12204230), the Guangdong Provincial Key Laboratory (2019B121203002), the Pearl River Talent Recruitment Program (2019QN01X298), the Guangdong Provincial Quantum Science Strategic Initiative (GDZX2303001, GDZX2200001) and the Science, Technology and Innovation Commission of Shenzhen Municipality (ZDSYS20190902092905285).
\end{acknowledgments}

%

\section{Appendix A: DMFT and pseudocode for the hybrid quantum-classical approach}
Dynamical mean field theory (DMFT) has demonstrated its success in studying a wide range of systems, particularly strongly correlated electron materials like transition metal oxides and heavy fermion compounds. It has provided valuable insights into phenomena such as metal-insulator transitions, Mott transitions, and the emergence of exotic phases.
However, the most computationally demanding aspect of DMFT lies in solving the impurity problem during each self-consistency iteration. This step involves addressing a quantum many-body problem for the impurity site, which is embedded in an effective bath. Numerical techniques, such as quantum Monte Carlo (QMC) or exact diagonalization (ED), are commonly employed to tackle the impurity problem.

The complexity of the impurity model increases exponentially with the number of orbitals, posing computational challenges. Dealing with multi-orbital systems requires extensive computational resources and efficient algorithms to handle the additional degrees of freedom and the interactions among different orbitals. As a result, accurately solving the impurity problem becomes the most demanding computational task within the DMFT framework.
To overcome the challenges posed by exponential complexity, we employ a hybrid quantum-classical algorithm. In this approach, the computationally intensive impurity problem is addressed using a quantum computer, while the computationally less demanding tasks are executed on a classical computer. This strategy allows us to leverage the strengths of both quantum and classical computing to optimize efficiency and mitigate the exponential complexities involved.

\begin{figure*}
  \centering
  \includegraphics[width=0.85\textwidth]{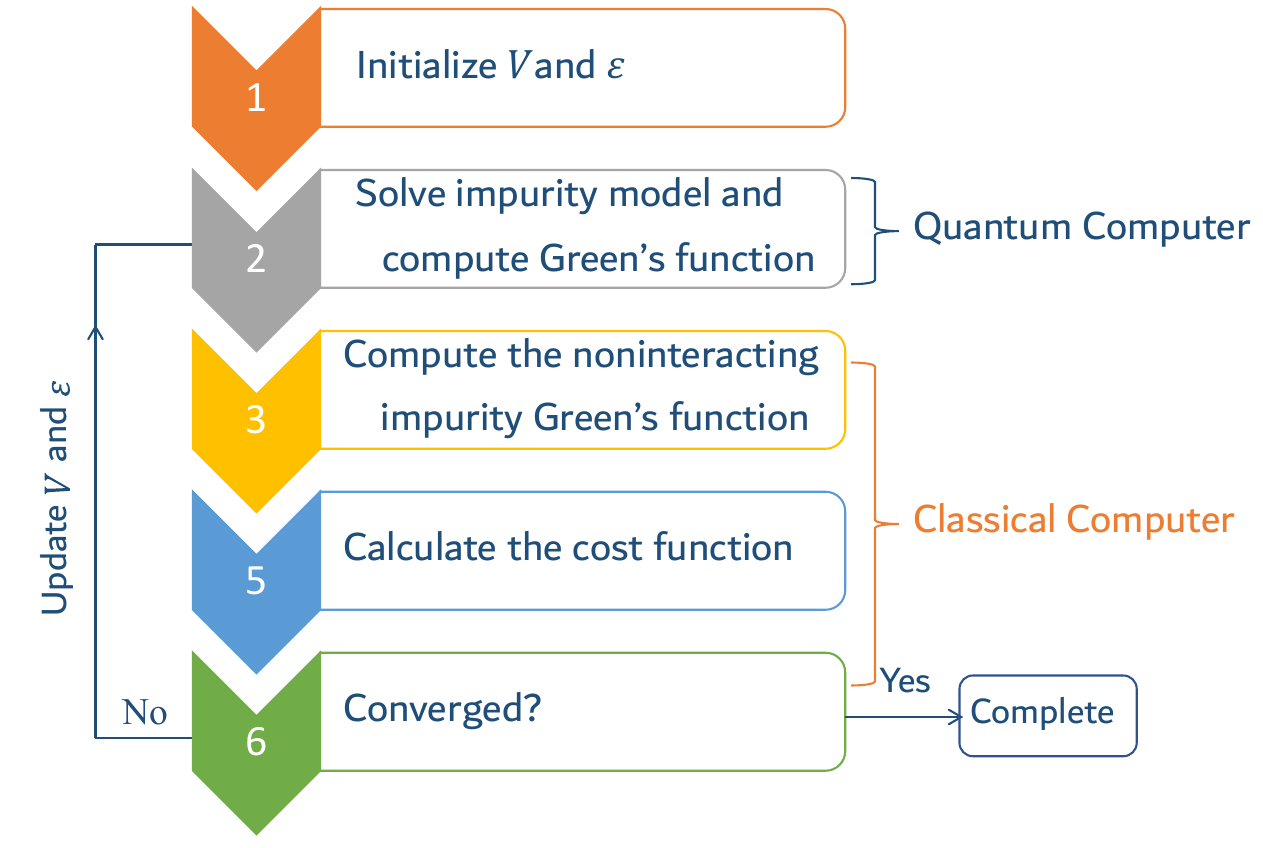}
  \caption{Flowchart for the DMFT calculation implemented on a hybrid quantum-classical system. The loop is iterated until two successive values of $V$ and $\varepsilon$ are within a specified threshold of each other.
  The impurity model solver and Green's function computation are executed on a quantum computer, while other tasks are performed on a classical computer.}\label{FigS1}
\end{figure*}

To minimize the numbers of qubit and the quantum gates required in the quantum simulation experiment, we firstly considered the single-impurity model with only one bath site, so that only one impurity-bath coupling $V$ and one bath energy $\epsilon$ have to be determined during the iterative process, as shown in Fig.~\ref{FigS1}. In the main text, we consider about the half-filled Hubbard model on a Bethe lattice, and use the single-impurity model with one bath site and fixed bath energy (in this work, $\epsilon=0$) to approximate it. The pseudocode for solving such impurity problems is presented in Table~\ref{DMFT}.

\begin{table*}[h!]
\begin{tabular}{p{0.95\textwidth}}
\toprule
\textbf{Algorithm}  Hybrid quantum-classical algorithm to solve the single-impurity model with one bath site (bath energy $\epsilon$ fixed)\\[0.5pt]
\hline
\textbf{Input}: The value of $U$ and the initial guess of $V$\\[0.5pt]
\textbf{Output}: The final value of $V$ \\[0.5pt]
1. With each iteration, measure correlation  \\
\quad\quad\quad $ O_1(t)$ and $ O_2(t)$ \\
\quad  on a quantum computer.\\
2. Compute the particle and whole interacting Green's function\\
 \quad   $G^p(t)$ and  $G^h(t)$ on a classical computer.   \\[0.5pt]

3. Compute the interacting Green's function in imaginary frequencies: \\
\quad\quad\quad $G(i\omega_n)=-i\int_{\epsilon}^\infty dte^{-t(\eta+\omega_n)\left[G^p(t)+\bar{G}^{h}(t)\right]}$.\\[0.5pt]

4. Compute the cost function by\\
\quad\quad\quad $f(V)=\sum_n\left\vert \tilde{G}_0^{-1}(i\omega_n)-G_0^{-1}(i\omega_n)\right\vert^2$ \\
\quad\quad\quad with $\tilde{G}_0^{-1}(i\omega_n)=i\omega_n+\mu-G(i\omega_n)$ and
$ G_0^{-1}(i\omega_n)=i\omega_n+\mu-\frac{V^2}{i\omega_n-\epsilon}$,\\
\quad\quad\quad where $ G_0(i\omega_n)$ is the exact noninteracting Green's function determined by $V$ and $\epsilon$.\\

5. Change $V+\delta V \leftarrow V$, and repeat step 1 to 4 to obtain $f(V+\delta V)$.\\
\quad\quad\quad Update $(V-g\delta V)\longleftarrow V$, with $g=\frac{f(V+\delta V)-f(V)}{\delta V}$.\\

6. Repeat step 1 to 5. \\
\quad End and output $V$ if $f$ converges.\\

7. Compute the density of states by\\
\quad $A(\omega)=-\text{Im} G(\omega)/\pi$.\\

\toprule
\end{tabular}
\caption{Pseudocode of the hybrid quantum-classical algorithm to solve the single-impurity model with one bath site (bath energy $\epsilon$ fixed) for approximating the Hubbard model embedded on a Bethe lattice.}
\label{DMFT}
\end{table*}

\section{Appendix B: SIAM model and Green's function}
A well-known model for describing strongly correlated systems is the Hubbard model, defined by the following Hamiltonian:
\begin{equation}\label{EqS1}
    \mathcal{H}= -t\sum_{<i,j>\sigma}(\hat{c}_{i\sigma}^{\dagger}\hat{c}_{j\sigma}+h.c.)+U\sum_i\hat{n}_{i,\downarrow}\hat{n}_{i,\uparrow}.
\end{equation}

To tackle this model, we employ the DMFT approach, which simplifies the treatment of spatial fluctuations around a single lattice site and approximates the effect of the many-body lattice through a self-consistent mean-field. Simultaneously, the isolated lattice site, which retains the on-site interaction, dynamically interacts with this mean-field, justifying the term DMFT. This approach significantly reduces the complexity of the original Hubbard model and approaches exactness as the dimension of the bath approaches infinity. To represent DMFT, we use the single-impurity Anderson model (SIAM), with one fermionic site as the impurity and another fermionic site as the bath, simulating the mean-field as follows:
\begin{equation}\label{EqS2}
\begin{aligned}
    \mathcal{H}_{\text {SIAM }}=U \hat{n}_{1 \downarrow} \hat{n}_{1 \uparrow}-\mu \sum_{\sigma} \hat{n}_{1 \sigma}+
    \sum_{\sigma} \epsilon~\hat{c}_{2 \sigma}^{\dagger} \hat{c}_{2 \sigma}+\\ \sum_{\sigma} V(\hat{c}_{1 \sigma}^{\dagger} \hat{c}_{2 \sigma}+ h.c.).
\end{aligned}
\end{equation}
Here, $U$ represents the Hubbard interaction at the impurity site 1, and $\mu$ is the chemical potential at that site. In this work, we consider about the half-filled Hubbard model, i.e., $\mu = U/2 $. Parameters $\epsilon$ and $V$ are determined through a self-consistent loop to match the original Hubbard model. The matching condition is given by:
\begin{equation}\label{EqS3}
    G\left(i \omega_{n}\right)=\int_{-\infty}^{\infty} dx \frac{D(x)}{i \omega_{n}+\mu-\Sigma\left(i \omega_{n}\right)-x},
\end{equation}
where $G(i\omega_n)$ is the Green's function in discrete imaginary frequencies for SIAM, $D(\epsilon)$ represents the density of states, and $\Sigma$ is the self-energy, calculated as:
\begin{equation}\label{EqS4}
    \Sigma(i\omega_n)=G_0^{-1}(i\omega_n)-G^{-1}(i\omega_n).
\end{equation}
Here, $G_0$ denotes the non-interacting impurity Green's function.

Calculating Green's functions for quantum impurity models is a central task that hinges on two critical steps: obtaining the ground state of the impurity model and simulating its dynamical evolution. Generally, estimating the ground energy or obtaining the ground state of a quantum many-body system, composed of either spins or fermionic modes with local interactions, is a non-trivial problem. It is formally classified as QMA-complete, reflecting its computational difficulty [Nat. Phys. 5, 732 (2009)].
For quantum impurity problems specifically, the best-known classical algorithm for estimating the ground energy of such a model has been outlined in previous research [Commun. Math. Phys. 356, 451 (2017)]. This algorithm shows that the ground energy of an impurity model, characterized by an impurity size $m$ and a bath size $n$, can be approximated within a precision $\gamma$ in time $O(n^3)\exp[O(m\log^3(m\gamma-1)]$. To achieve this approximation, a low-energy state is constructed as a superposition of fermionic Gaussian states to approximate the ground state. The complexity of computing ground states for quantum impurity problems increases substantially with the impurity size, as indicated by this time complexity.
Regarding the time evolution, in the above paper, researchers has also discovered that classical algorithms cannot efficiently simulate the time-dependent impurity Hamiltonian. This was proven by investigating a special impurity Hamiltonian detailed in [Quantum Inf. Comput. 14, 901 (2014)]. However, the efficiency of simulating time-independent Hamiltonians remains an open question.
As a result, the computational effort required to determine the Green's function increases significantly with the size of the impurity, underscoring the escalating challenge faced in such calculations.

Return to our experiment, to simplify Eq.~(\ref{EqS4}), let us consider the Hubbard model on a Bethe lattice and set the coordination number to infinity (each site connects with an infinite number of sites). The density of states then becomes:
\begin{equation}\label{EqS5}
    D(x)=\frac{\sqrt{4t^{\star2}-x^2}}{2\pi t^{\star^2}}
\end{equation}
and the hopping term $t$ in the original Hubbard model becomes:
\begin{equation}\label{EqS6}
    t^{\star}=\frac{t}{\sqrt{z}},\quad z\rightarrow\infty
\end{equation}
In this work, we use the value $t^{\star}=1$.

In this specific case, the integral in Eq.~(\ref{EqS3}) simplifies to:
\begin{equation}\label{EqS7}
    G_{0}^{-1}\left(i \omega_{n}\right)=i \omega_{n}+\mu-G\left(i \omega_{n}\right)
\end{equation}
This implies that we can obtain an approximate non-interacting Green's function $\tilde{G}_0$ using the impurity Green's function $G$. In this work, for the specific case of a discrete bath, the exact non-interacting Green's function can be given as:
\begin{equation}\label{EqS8}
    G^{-1}_{0}\left(i \omega_{n}\right)=i \omega_{n}+\mu-\frac{V^{2}}{i \omega_{n}-\epsilon}.
\end{equation}
Consequently, we can define the cost function $f$ as follows:
\begin{equation}\label{EqS9}
    f=\sum_{n}|\tilde{G}^{-1}_{0}(i \omega_{n})-G^{-1}_{0}(i \omega_{n})|
\end{equation}
to update the parameters $\epsilon$ and $V$.

To obtain the impurity Green's function in imaginary frequencies, we define the particle and hole Green's functions of the SIAM as:
\begin{equation}\label{EqS10}
    \begin{aligned}
        G^{\mathrm{p}}(t) &=\langle\Psi|c_{1}(t) c_{1}^{\dagger}(0)| \Psi\rangle \\
        G^{\mathrm{h}}(t) &=\langle\Psi|c_{1}^{\dagger}(t) c_{1}(0)| \Psi\rangle.
    \end{aligned}
\end{equation}
Here, $|\Psi\rangle$ is the ground state of SIAM, $c_1$ and $c^{\dagger}_1$ denote the annihilation and creation operators on the impurity site 1. These functions can be computed from $O_1(t)$ and $O_2(t)$, which are defined as:
\begin{equation}\label{EqS11}
    \begin{aligned}
        O_1(t)& =\langle \Psi\vert e^{i\mathcal{H}t}q_1e^{-i\mathcal{H}t}q_1\vert\Psi\rangle,\\
        O_2(t)& =\langle \Psi\vert e^{i\mathcal{H}t}q_1e^{-i\mathcal{H}t}q_2\vert\Psi\rangle,
    \end{aligned}
\end{equation}
where $q_1=c_1+c_1^\dagger$ and $q_2=i(c_1-c_1^\dagger)$. In this case, we assume the absence of superconductivity (which holds in this specific case), so operators that do not conserve particle number won't contribute to the expectation values, maintaining particle-hole symmetry. The resulting particle and hole Green's functions are
\begin{equation}\label{EqS12}
    G^\text{p,h}(t)=\frac{1}{2}(O_1(t) \pm  O_2(t)  ).
\end{equation}

From these Green's functions, the spectral function $A(\omega)$ can be calculated as
\begin{equation}\label{EqS13}
    A(\omega)=-\frac{\text{Im}G(\omega)}{\pi}.
\end{equation}

\section{Appendix C: Jordan-Wigner transformation of the SIAM}
To implement the two-site SIAM with nuclear qubits, we need to map the fermionic creation and annihilation operators onto tensor products of spin operators. These spin operators then act on the qubits through quantum gates. To simplify the quantum gates, we adopt a specific qubit ordering. In this ordering, the first two qubits represent the spin-up state for both fermionic sites, while the last two qubits represent spin-down. This qubit ordering is achieved through the Jordan-Wigner transformation, a mathematical mapping connecting fermionic operators with spin operators. By explicitly applying this transformation, we establish a correspondence between fermionic degrees of freedom and qubit spin states. This enables us to manipulate the qubits with quantum gates to simulate the desired fermionic system. By utilizing the Jordan-Wigner transformation and suitable quantum gates, we can effectively encode and manipulate fermionic states within the qubit framework, enabling the simulation of the two-site SIAM. This approach offers a way to study and analyze fermionic systems using quantum computing techniques.

For the two-site SIAM we studied here, the Jordan-Wigner transformation works as follows:
\begin{align}\label{EqS11}
\hat{c}^{\dagger}_{1\downarrow} &= \sigma^{-}_{1} = \sigma_{x}^{1} - i\sigma_{y}^{1}, \nonumber\\
\hat{c}^{\dagger}_{2\downarrow} &= \sigma_{z}^{1} \sigma^{-}_{2} = \sigma_{z}^{1}(\sigma_{x}^{2} - i\sigma_{y}^{2}), \\
\hat{c}^{\dagger}_{1\uparrow} &= \sigma_{z}^{1}\sigma_{z}^{2} \sigma^{-}_{3} = \sigma_{z}^{1}\sigma_{z}^{2}(\sigma_{x}^{3} - i\sigma_{y}^{3}),\nonumber\\
\hat{c}^{\dagger}_{2\uparrow} &= \sigma_{z}^{1}\sigma_{z}^{2}\sigma_{z}^{3} \sigma^{-}_{4} = \sigma_{z}^{1}\sigma_{z}^{2}\sigma_{z}^{3}(\sigma_{x}^{4} - i\sigma_{y}^{4}),\nonumber
\end{align}
where $\hat{c}_{i\sigma}=(\hat{c}^{\dagger}_{i\sigma})^{\dagger}$, and $\sigma_{j,\alpha}$ is the Pauli matrix of the $j$-th spin in the $\alpha$ direction. For a larger spin system, the Jordan-Wigner transformation is in the form of $\hat{c}^{\dagger}_{j\downarrow}=\left(\prod_{p<2j-1}\sigma_z^p\right)\sigma^{-}_{2j-1}$
and $\hat{c}^{\dagger}_{j\uparrow}=\left(\prod_{p<2j+1}\sigma_z^p\right)\sigma^{-}_{2j+1}$.

Thus, the half-filled SIAM can be written as the tensor product of spin operators as
\begin{equation}\label{EqS12}
\mathcal{H}=\frac{U}{4}\sigma_z^1\sigma_z^3+\frac{V}{2}(\sigma^1_x\sigma^2_x+\sigma^1_y\sigma^2_y+\sigma^3_x\sigma^4_x+\sigma^3_y\sigma^4_y)
\end{equation}
with the mappings in Eq.~(\ref{EqS11}), which can be conveniently realized with a quantum spin simulator. The number of qubits, denoted as $N$, required for the implementation scales linearly with the total number of sites in the impurity and the bath, i.e., $N=2(n_{i}+n_b)$, where $n_i$ and $n_b$ are the number of the impurity and bath sites, respectively. The correlation functions measured in experiments are given as:

\begin{equation}\label{EqS13}
\begin{aligned}
O_1(t)&=\langle \Psi\vert e^{i\mathcal{H}t}\sigma^1_xe^{-i\mathcal{H}t}\sigma^1_x\vert\Psi\rangle,\\
O_2(t) &=\langle \Psi\vert e^{i\mathcal{H}t}\sigma^1_xe^{-i\mathcal{H}t}(-\sigma^2_y)\vert\Psi\rangle.
\end{aligned}
\end{equation}

\section{Appendix D: NMR implementation}
\begin{figure*}
  \centering
  \includegraphics[width=0.8\textwidth]{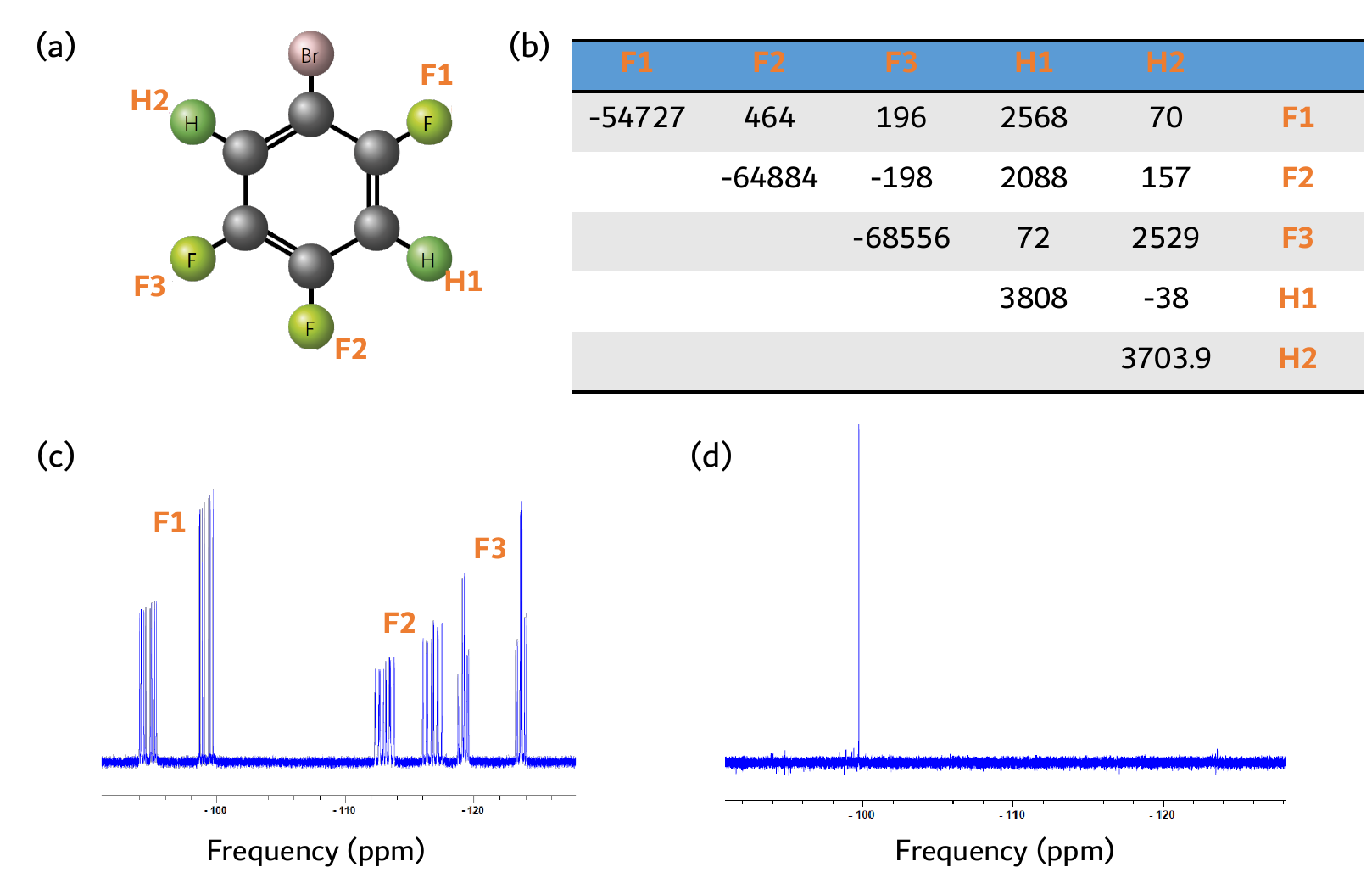}
  \caption{(a) Molecular structure of the sample.
  (b) Parameters of the sample in units of Hz.
  (c) Thermal spectrum of the fluorine nuclei.
  (d) Spectrum of the PPS.}\label{FigS2}
\end{figure*}

\textit{Sample.}---Our experiments utilized a five-qubit quantum processor, which is composed of two $^1$H nuclear spins and three $^{19}$F nuclear spins within 1-bromo-2,4,5-trifluorobenzene dissolved in liquid crystal solvent MBBA. The structure of  this compound is shown in Fig.~\ref{FigS2}(a). F1, F2, F3, H1, and H2 spins were designated as qubits 1 through 5, respectively. 
The experiments were carried out using a Bruker 600 MHz spectrometer at $305$ K.

The natural Hamiltonian of this quantum system is described by the following equation
\begin{equation}\label{EqS14}
\begin{aligned}
  \mathcal{H} = \sum\limits_{j=1}^5 \pi \nu_j \sigma_z^j + \sum\limits_{j < k,=1}^5 \frac{\pi}{2}( J_{jk}+2D_{jk}) \sigma_z^j \sigma_z^k.
\end{aligned}
\end{equation}
In this equation, $\sigma_z^j$ represents the Pauli-Z matrix of the $j$-th spin, $\nu_j$ denotes the chemical shifts of the $j$-th spin, and $(J_{jk}+2D_{jk})$ represents the effective coupling strengths, as shown in Fig.~\ref{FigS2}(b).
We applied the secular approximation due to the significant difference between the chemical shift in each pair of spins and the effective coupling strength. The thermal equilibrium NMR spectrum of the sample are illustrated in Fig.~\ref{FigS2}(c), respectively. An imbalance is observed specifically in the peak heights across the spectrum, while the integral values (signal intensities) of the peaks maintain consistency across the spectrum.
The variation in peak heights across the multiplets of the $^{19}$F nuclei is caused by the differences in relaxation timescales among the peaks within each multiplet. Specifically, the disparity among peaks from different nuclei is largely due to the differential transverse relaxation times ($T_2$) of the $^{19}$F nuclei. For instance, F1 exhibits the longest $T_2$ among the three. Furthermore, the observed imbalance within multiplets of the same nucleus, especially the signal on the right-hand side of each multiplet, is attributed to cross-correlation effects. 

\textit{Adiabatic initialization.}---In the experiment, we employed a scattering circuit to measure the Green's function, requiring a total of five qubits. In our experimental setup, the first spin, F$_1$, was designated as the ancillary qubit, while the other four spins were utilized to simulate the two-site impurity model.

To initialize the spin system for the DMFT algorithm, we began with the PPS state (the preparation of which is detailed in the next section) and set the system in the initial state $\rho_0 = \vert+\rangle \otimes \vert\psi_g\rangle$. The ancillary qubit could be easily prepared in the superposition state $\vert+\rangle$ using the Hadamard gate. However, preparing the other spins in the state $\vert \psi_g\rangle$ was a highly challenging task in our experiment.

To obtain the ground state, we employed a combination of adiabatic state preparation (ASP). The procedure started with the preparation of the ground state of a simple Hamiltonian, $\mathcal{H}_0$, which is relatively easy to achieve. Subsequently, the system was evolved under a time-varying Hamiltonian $\mathcal{H}(t)$ that smoothly transitioned from $\mathcal{H}_0$ to the desired Hamiltonian described by Eq.~(\ref{EqS5}). The key aspect was to change the Hamiltonian's parameters slowly, ensuring that the evolution occurred at a rate much slower than the inverse spectral gap of $\mathcal{H}(t)$. This approach guaranteed that the wave function of the system remained close to the ground state throughout the evolution process.

The initial Hamiltonian chosen for the experiment was $\mathcal{H}_0 = \frac{V}{2}(\sigma^1_x\sigma^2_x+\sigma^1_y\sigma^2_y+\sigma^3_x\sigma^4_x+\sigma^3_y\sigma^4_y)$. The corresponding ground state was $\vert\Psi_0\rangle = \frac{1}{2}(\vert0101\rangle-\vert0110\rangle-\vert1001\rangle+\vert1010\rangle)$ provided $V\geq0$ is satisfied. $\vert\Psi_0\rangle$ could be prepared from $\vert 0000\rangle$ through three operations as follows:
\begin{align}\label{EqS18}
  U_1 &= e^{-i\sigma_y^2\pi/4}e^{-i\sigma_y^4\pi/4},\nonumber \\
  U_2 &= \sigma_x^1\otimes\vert00\rangle\langle00\vert\otimes \sigma_x^4\nonumber \\
  &+ I_2\otimes(\vert01\rangle\langle01\vert+\vert10\rangle\langle10\vert+\vert11\rangle\langle11\vert)\otimes I_2, \\
  U_3 &= (\vert01\rangle\langle01\vert+\vert10\rangle\langle10\vert)\otimes e^{i\sigma_y^3\pi/4}e^{i\sigma_y^4\pi/4}\nonumber \\
  &+(\vert00\rangle\langle00\vert+\vert11\rangle\langle11\vert)\otimes I_4.\nonumber
\end{align}
Here, $I_2$ and $I_4$ represent $2\times2$ and $4\times4$ identity matrices, respectively.

\begin{figure*}
  \centering
  \includegraphics[width=0.8\textwidth]{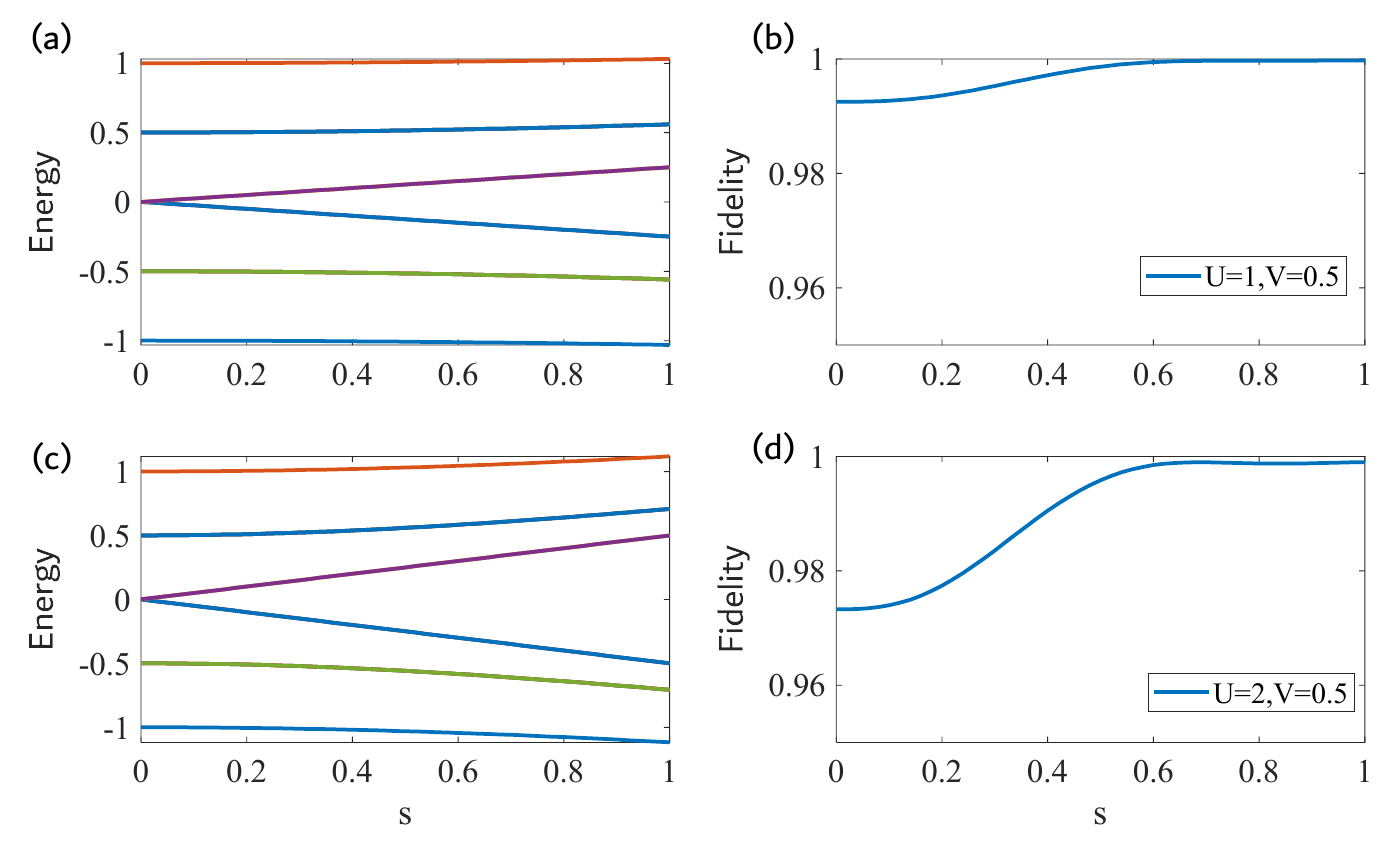}
  \caption{
Energy structure and fidelity of ASP were assessed as functions of $s$ for different values of $V$ and $U$ within a single run. Figures (a) and (c) represent the energy structures for $U=1.0, V=0.5$ and $U=2.0, V=0.5$, respectively, while Figures (b) and (d) display the corresponding fidelities.
According to the adiabatic theorem, fidelity tends to approach 1 as the evolution time increases. In our numerical simulations, the adiabatic evolution was discretized into 50 steps, with a total time set to $T=4$. This analysis demonstrates that improving the fidelity of ASP can be achieved by extending the evolution time, aligning with the principles of adiabatic quantum computing.}\label{FigS5}
\end{figure*}

Once the ground state of $\mathcal{H}_0$ is obtained, we utilize the ASP technique to prepare the system for the ground state of $\mathcal{H}$ in Eq.~(\ref{EqS5}) with an arbitrary value of $U$. It is important to note that we investigate the properties of three different many-body interactions experimentally, i.e., $U=0, 1, 2$. In the case of $U=0$, the ground state of $\mathcal{H}$ always remains in $\vert\Psi_0\rangle$, so the ASP is not employed in this special scenario.

Practically, we transitioned the system's Hamiltonian slowly from $\mathcal{H}_0$ to $\mathcal{H}$ using linear interpolation, given by $\mathcal{H}_\text{ad} = (1 - s)\mathcal{H}_0 + s\mathcal{H}$, where $s = t/T$ and the time $t$ varies from $0$ to $T$. As $T$ increases, the state of the system qubit approaches the desired state $\vert\Psi\rangle$ through ASP. To assess the fidelity of ASP, defined as $\vert \langle\Psi\vert\Psi_0\rangle\vert^2$, we conducted numerical simulations and examined its dependence on $s$, as shown in Fig.~\ref{FigS5}. In our numerical simulations, we selected $T=4$, and the evolution is divided into $M=50$ steps.

The unitary evolution for each adiabatic step is denoted as $u_m = e^{\mathcal{H}_\text{ad}(m)\Delta T}$, where
$\mathcal{H}_\text{ad}(m) = (1 - \frac{m}{M})\frac{U}{4}\sigma_z^1\sigma_z^3 + \frac{V}{2}(\sigma^1_x\sigma^2_x + \sigma^1_y\sigma^2_y + \sigma^3_x\sigma^4_x + \sigma^3_y\sigma^4_y)$, and $\Delta T = T/M$.

The evolution $u_m$ can be decomposed into two-body coupling evolutions using the Trotter-Suzuki approximation, as follows
\begin{equation}\label{EqS19}
\begin{split}
  u_m \simeq & \, e^{-i\left(1-\frac{m}{M}\right)\sigma_z^1\sigma_z^3 U \Delta T / 4} \\
  & e^{-i\sigma_x^1\sigma_x^2 V \Delta T / 2} \\
  & e^{-i\sigma_y^1\sigma_y^2 V \Delta T / 2} \\
  & e^{-i\sigma_x^3\sigma_x^4 V \Delta T / 2} \\
  & e^{-i\sigma_y^3\sigma_y^4 V \Delta T / 2}.
\end{split}
\end{equation}
This decomposition can be conveniently implemented on a quantum simulator.

\begin{figure}
  \centering
  \includegraphics[width=0.48\textwidth]{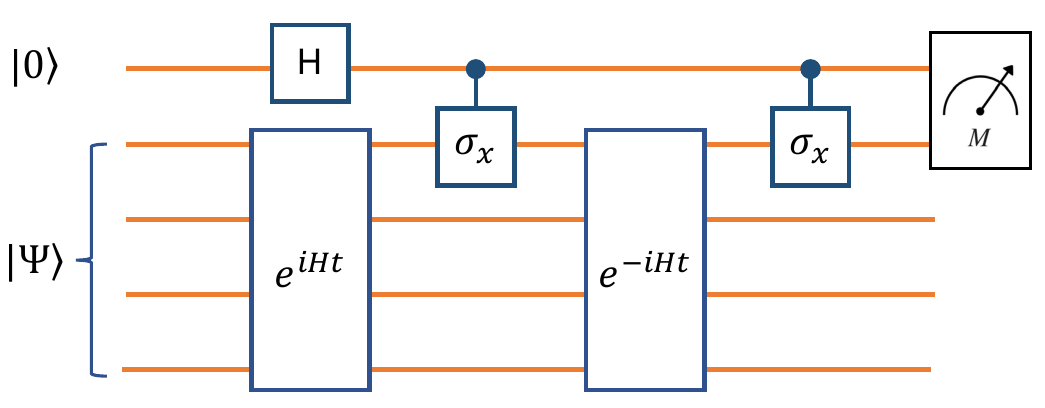}
  \caption{Quantum scattering circuit to measure the correlation function
  $\langle \Psi\vert e^{i\mathcal{H}t}\sigma_x^1e^{-i\mathcal{H}t}\sigma_x^1\vert\Psi\rangle $,
while $\langle \Psi\vert e^{i\mathcal{H}t}\sigma_x^1e^{-i\mathcal{H}t}(-\sigma_y^1)\vert\Psi\rangle$ can be measured by replacing the second control-$\sigma_x^1$ with a control-$\sigma_y^1$ operation.}\label{FigS6}
\end{figure}

\textit{Scattering circuit.}---As mentioned in Eq.~(\ref{EqS13}), the observable measured experimentally is the correlation function, which can be viewed as the overlap of two quantum states $\vert\Psi\rangle$ and $U_{1,2}(t)\vert\Psi\rangle$, with $U_1(t)=e^{i\mathcal{H}t}\sigma_x^1e^{-i\mathcal{H}t}\sigma_x^1$ and $U_2(t)=e^{i\mathcal{H}t}\sigma_x^1e^{-i\mathcal{H}t}(-\sigma_y^1)$ .
Taking the time correlation function $\langle \Psi\vert U_1(t)\vert\Psi\rangle$ for example, it can be measured via the scattering circuit presented in Fig.~\ref{FigS6}.
In a scattering circuit, a probe qubit is deliberately placed in a known initial state and interacts with the system. This interaction is designed in such a way that a subsequent measurement of the probe qubit's state provides valuable information about the state of the system.
The quantum circuit works as follows.
Considering the input state $\rho_\text{probe}\otimes\rho_\text{system}=\vert0\rangle\langle0\vert\otimes\vert\Psi\rangle\langle\Psi\vert$,
where the first qubit on the left is the probe, entering the circuit on $\vert0\rangle$, and the other one is the
system, prepared in the ground state of the SIAM $\vert\Psi\rangle$.
Upon the transformation shown in Fig.~\ref{FigS4}, the final state of total system terms into
$\rho_t=\vert\psi_t\rangle\langle\psi_t\vert$,
with
\begin{equation}\label{EqS20}
  \vert\psi_t\rangle=\frac{1}{\sqrt{2}}\left(\vert0\rangle\otimes\vert\Psi\rangle+\vert1\rangle\otimes U_1(t)\vert\Psi\rangle\right).
\end{equation}
The time correlation function is then extracted as a non-diagonal operator of the probe qubit, $2\text{Tr}(\vert1\rangle\langle0\vert\rho_t)$.
We further recall that $\vert1\rangle\langle0\vert=(\sigma_x-i\sigma_y)/2$,
so that the time correlation is corresponds to
\begin{equation}\label{EqS21}
  \langle\Psi\vert U_1(t)\vert\Psi\rangle={\langle\sigma_x\rangle-i\langle\sigma_y\rangle}.
\end{equation}
In NMR system, $\langle\sigma_x\rangle$ and $\langle\sigma_y\rangle$  correspond to the transverse magnetization and can be readout directly.
The correlation function $\langle\Psi\vert U_2(t)\vert\Psi\rangle$ can also be measured with the similar operation by
substituting the last control-$\sigma_x^1$ with a control-$\sigma_y^1$.
To improve the precision of the experiment, the circuit shown in Fig.~\ref{FigS4} is realized by a shape pulse search by a optimization algorithm with a length of $40$ ms.

\textit{Implementation of the evolution $e^{-i\mathcal{H}t}$.}---The evolution $e^{-i\mathcal{H}t}$ can be realized by the Trotter-Suzuki formula
\begin{equation}
  e^{-i\mathcal{H}t}\approx [e^{-i\mathcal{H}_z^{13}\theta_1}e^{-i\mathcal{H}_x^{12}\theta_2}e^{-i\mathcal{H}_y^{12}\theta_2}
  e^{-i\mathcal{H}_x^{34}\theta_2}e^{-i\mathcal{H}_y^{34}\theta_2} ]^n,
\end{equation}
where the evolution time $t$ is divided into $n$ segments, with $\theta_1=Ut/4n$, $\theta_2=Vt/2n$, and $\mathcal{H}_\alpha^{ij}=\sigma_\alpha^i\sigma_\alpha^j$ representing the two-body coupling Hamiltonian. In experiment, we consider the first-order Trotter-Suzuki approximation for $n=1$. The controlled-$e^{-i\mathcal{H}_\alpha^{ij}\theta}$ gate is achieved through optimized radio-frequency pulses combined with the NMR refocusing technique, 
where the relevant pulse sequence is depicted in Fig.~2(b) of the main text. 

\textit{Measurement.}---In a five-qubit NMR quantum processor, the signal from each spin is typically split into $16$ peaks, which arise from the couplings between different nuclei. In NMR spin dynamics, these peaks consist of both real and imaginary parts. The real part encodes the expectation values of the observable spin's Pauli matrix $\sigma_x$, while the imaginary part encodes the expectation values of $\sigma_y$. The remaining spins are measured in the computational basis.
Considering the probe qubit in our experiment, the peaks correspond to the other four qubits being in the states $\left\{0000, 0001, 0010, \cdots, 1111\right\}$.
We can express $\vert0\rangle\langle0\vert$ as $\left(I+\sigma_z\right)/2$ and $\vert1\rangle\langle1\vert$ as $\left(I-\sigma_z\right)/2$. This allows us to measure the expectation values of single-quantum coherence operators involving $\sigma_x$ or $\sigma_y$ in the target qubit, and $\sigma_z$ or $I$ in the remaining qubits.
To measure $\langle\sigma_x\rangle$ in the probe qubit, we can read out the observable $\sigma_xIIII$. It is important to note that $IIII = \vert0000\rangle\langle0000\vert + \vert0001\rangle\langle0001\vert + \cdots + \vert1111\rangle\langle1111\vert$. Therefore, $\langle\sigma_x\rangle$ represents the sum of the real part of all the peaks of the probe qubit. Similarly, $\langle\sigma_y\rangle$ represents the sum of the imaginary part of all the peaks of the probe qubit.

\section{Appendix E: Pseudo-pure state preparation}
\textit{Line-selective PPS preparation.}--
In our study, we adopted a line-selective approach, inspired by the pioneering research of X. Peng et al. [Chem. Phys. Lett. 340, 509 (2001)]. Building upon the foundational method, our work features an advanced design where the rotation angles for the line-selective pulses are analytically derived, negating the necessity for numerical optimization and marking a significant advancement from the original approach. Additionally, our technique avoids the necessity for a precisely designed shaped pulse to remove the zeroth-order quantum coherences.

\begin{figure}
  \centering
  \includegraphics[width=0.5\textwidth]{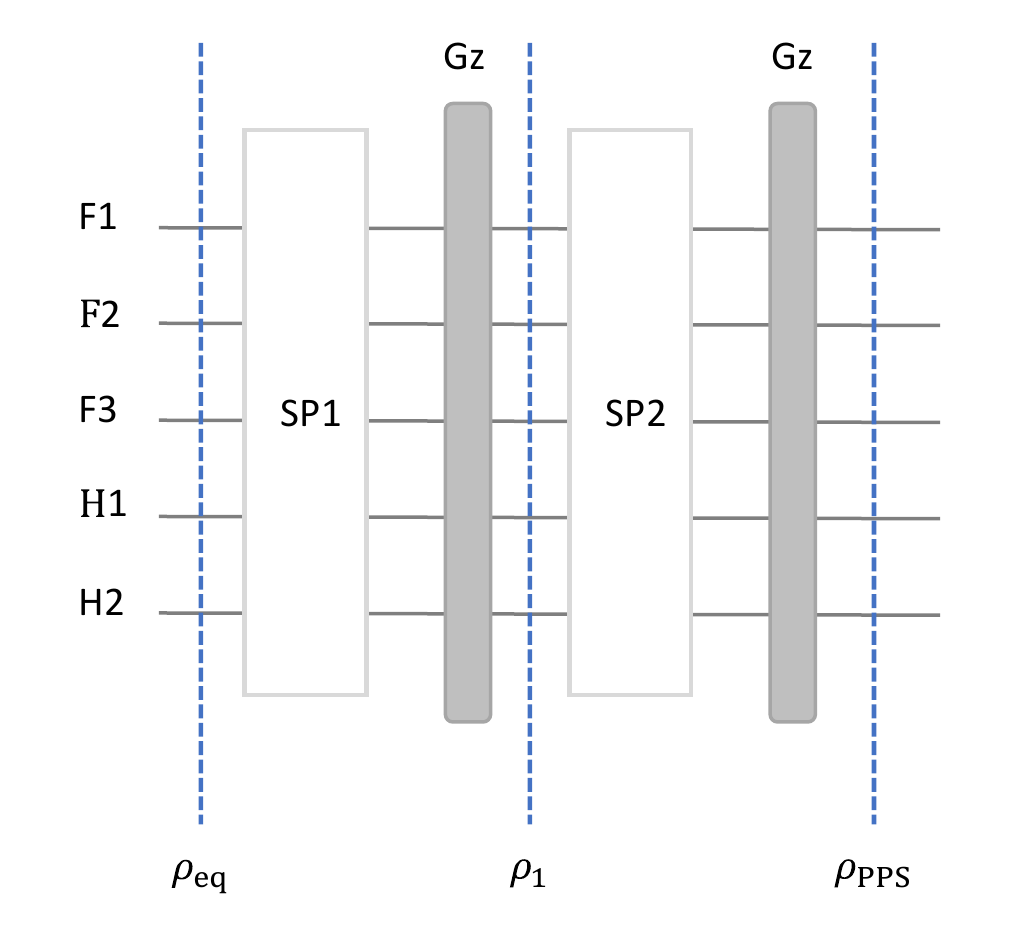}
  \caption{The quantum circuit to prepare the five-qubit system from the thermal equilibrium state to PPS, which composes of two shaped pulses and two z-gradient pulses.}\label{FigS3}
\end{figure}

The pulse sequence to prepare the PPS constitutes two line-selective pulses and two gradient pulses as shown in Fig.~\ref{FigS3}. The primary objective of the first shaped pulse is to selectively deplete the population of all other energy levels, leaving only the two targeted states, $\vert00000\rangle$ and $\vert11111\rangle$, with significant populations. The second pulse is employed to redistribute the population from the $\vert11111\rangle$ state to other states, leaving the state $\vert00000\rangle$ unaffected. We will detail our method in the following. 

\begin{figure*}
  \centering
  \includegraphics[width=0.9\textwidth]{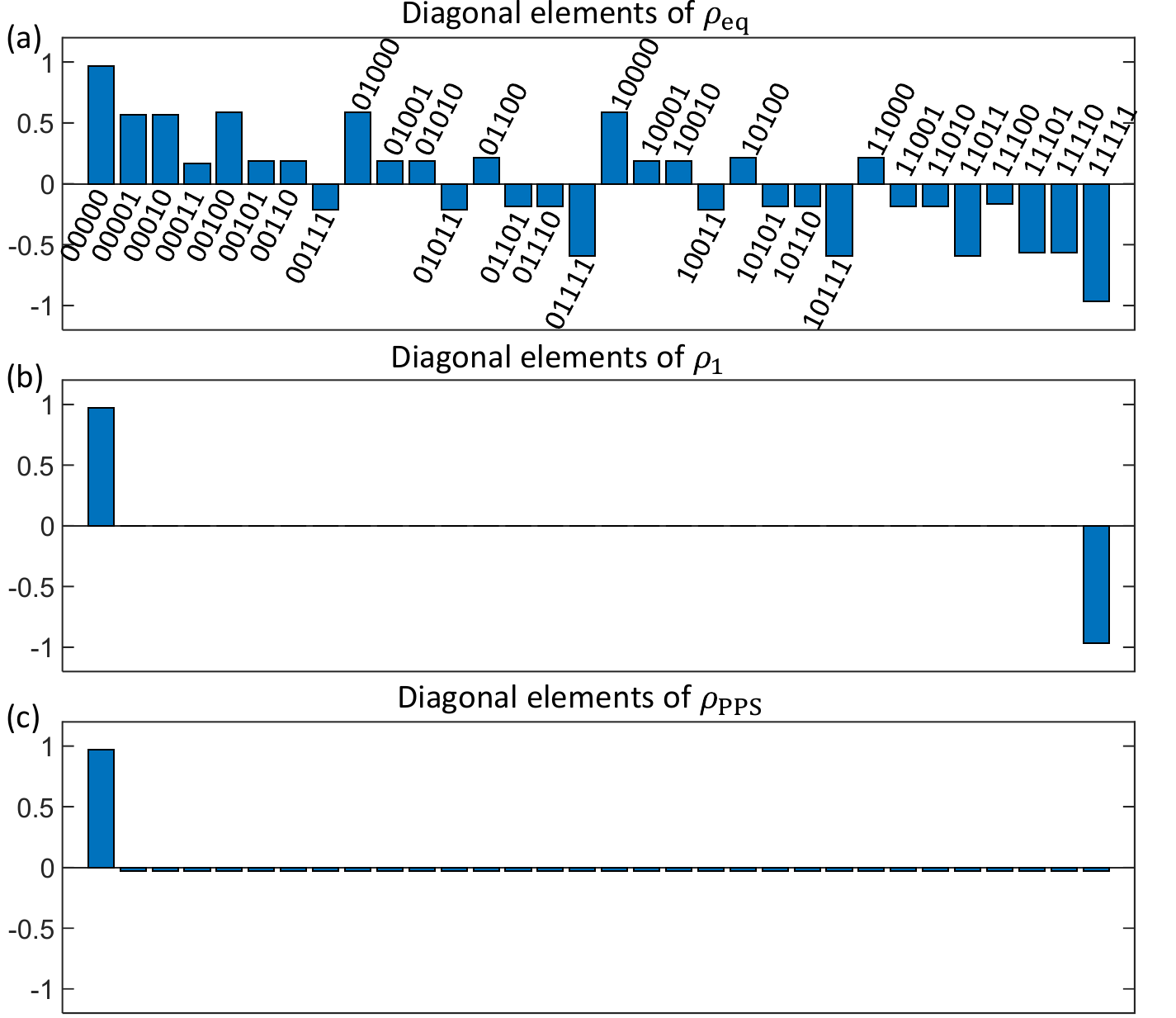}
  \caption{Evolution of the five-qubit system during the PPS preparation process, depicted through the diagonal elements of the density matrices at key stages: the thermal equilibrium state ($\rho_\text{eq}$), the state post-first z-gradient pulse ($\rho_1$) and the PPS ($\rho_\text{PPS}$) utilizing the line-selective method.}\label{FigS4}
\end{figure*}

All experiments were performed on a Bruker Ascend 600 MHz spectrometer (14.1 T) at $305$ K. The thermal equilibrium state of the five-qubit system is a highly mixed state. For the large identity matrix does not evolve under any unitary propagator and cannot be observed in NMR,  the thermal equilibrium state can be described by a deviation density matrix in the form of 
\begin{equation}\label{EqS15}
 \rho_\text{eq}=\gamma_\text{F}/\gamma_\text{H}(\sigma_z^1+\sigma_z^2+\sigma_z^3)+(\sigma_z^4+\sigma_z^5).
\end{equation}
Here, $\gamma_\text{H}$ and $\gamma_\text{F}$ represent the gyromagnetic ratio of nucleus $^{1}$H and $^{19}$F, respectively, and $\sigma_z^{i}$ denotes the Pauli matrix along the z-axis for the $i$-th qubit. $\rho_\text{eq}$ is a sparse matrix with only nonzero diagonal elements, the specific diagonal values of which are shown in Fig.~\ref{FigS4}(a).

We have excluded the standard normalization condition in Eq.~\eqref{EqS15}, so $\text{Tr}(\rho_\text{eq})=0$ is fulfilled. Furthermore, it is noteworthy that the (relative) population of the state $\vert00000\rangle$ is the maximal one among all the computational basis labeled from 00000 to 11111, and the populations of a pair of symmetric states (taking 01010 and 10101 for example) are opposite, so that $P_{01010}+P_{10101}=0$ is satisfied. We leverage these properties to enhance the line-selective method.

The objective of the first shaped pulse is to selectively deplete the population of energy levels, leaving only the two targeted states, $\vert00000\rangle$  and $\vert11111\rangle$, with significant population. To achieve this goal, we employ a series of $\pi/2$ line-selective pulses, which are applied individually between each pair of symmetric states, excluding the states $\vert00000\rangle$  and $\vert11111\rangle$. These $\pi/2$ line-selective pulses effectively normalize the population of corresponding states by leveraging the inherent property of opposite populations in symmetric states. These line-selective pulses are packed into a large one and forms the first shaped pulse (SP1) in the pulse sequence. The coherence produced by SP1 are $\vert00001\rangle\langle11110\vert$,  $\vert00010\rangle\langle11101\vert$, $\cdots$, $\vert01111\rangle\langle10000\vert$ and their conjugates. There is no zeroth-order coherence produced, which simplifies the subsequent pulse design by eliminating the need for complex methods to remove zeroth-order coherence. This conclusion holds true for all nuclear systems with an odd number of qubits. After a gradient pulse, the density matrix turns to be diagonalized, denoted as $\rho_1$, with populations on the computational basis shown as Fig.~\ref{FigS4}(b).

The second shaped pulse (SP2) aims to redistribute the population on $\vert11111\rangle$ equally over all other states, excluding $\vert00000\rangle$. This is achieved through a series of line-selective rotations that connect the state $\vert11111\rangle$ with each other state. The rotations employ angles such as $2\sin^ {-1}\sqrt{1/31}$, $2\sin^{-1}\sqrt{1/30}$, $\cdots$, $2\sin^{-1}\sqrt{1/2}$. Those rotations sequentially transfer portions of $1/31$, $1/30$, $\cdots$, $1/2$ of the current population of $\vert11111\rangle$ to the respective states, so that each state holds a $1/31$ of the original population of $P_{11111}$ at the end of rotations. Finally, the goal of averaging populations on all other states except $\vert00000\rangle$ arrives. As the line selective operations occur between the all-spin-down state $\vert11111\rangle$ and other states with some-spin-down, zero coherence will not be present at the end of SP2. After a z-gradient pulse (Gz) to eliminate all the unwanted coherence, the PPS of this five-qubit system is obtained, with population distribution in Fig.~\ref{FigS4}(c).

Return to the experiment, we utilized the optimal-control algorithm to search for the two RF pulses. The shaped pulses used in the experiments had lengths of $30$ ms and  $25$ ms, respectively. All the pulses exhibited fidelities over $99.5\%$.
The spectrum of the PPS is also displayed in  Fig.~\ref{FigS2}(d).

\textit{Applicability of the method.}--
Although the number of qubits in our experimental system is odd, we must mention that the method works for all systems, regardless of whether the number of qubits is odd or even. At first glance, it appears that zeroth-order coherence terms produced during the SP1 step cannot be eliminated with a single gradient pulse, potentially compromising the method. However, upon further investigation, we found that the method is also effective for systems with an even number of qubits. We will elaborate on this conclusion in the next.

Without loss of generality, let us use the four-qubit (even number of qubits) system as the example. There are two types of nuclear systems with an even number of qubits: those with homogeneous nuclear spins and those with inhomogeneous nuclear spins. For the inhomogeneous case, we label the two different spins of the inhomogeneous nuclear system as A and B (with corresponding gyromagnetic ratio, $\gamma_A$ and $\gamma_B$), resulting two different structures of the inhomogeneous four-qubit system, namely ABBB and AABB. For uniformity, we can label the homogeneous case as structure AAAA. For clarity, we will consider a 4-qubit system in three scenarios. The zeroth-order coherence terms that may be produced by SP1 are located at the positions$\vert0011\rangle\langle1100\vert$, $\vert0101\rangle\langle1010\vert$, $\vert0110\rangle\langle1001\vert$, and their conjugates, resulting in a total of six non-zero zeroth coherence terms.

(a). Homogeneous nuclear systems (AAAA structure)

\begin{figure*}
  \centering
  \includegraphics[width=0.9\textwidth]{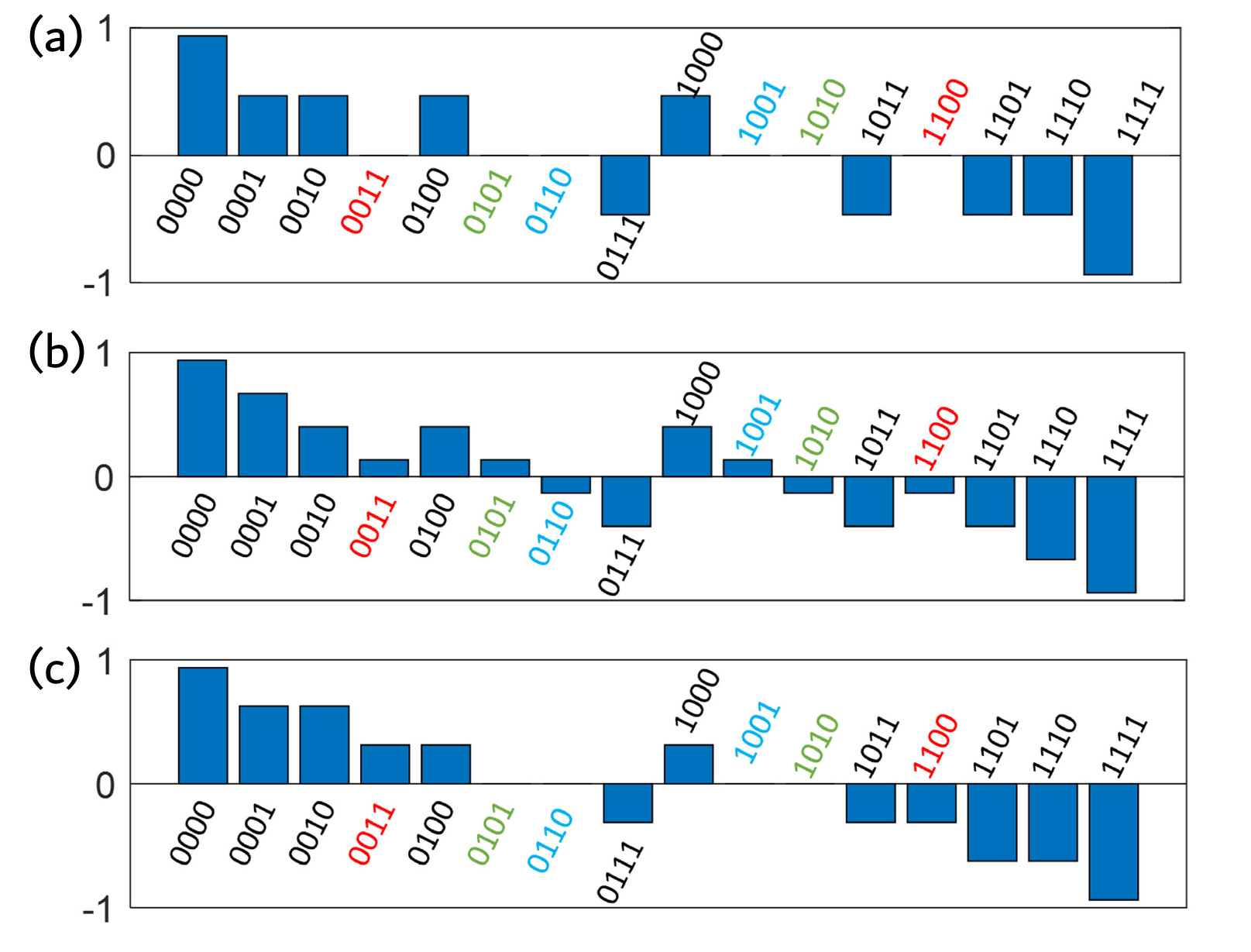}
  \caption{The corresponding diagonal elements of the thermalized equilibrium density matrix for the three structures when the nuclear system has four qubits: (a) Homogeneous nuclear systems (AAAA structure). (b) Inhomogeneous nuclear systems (ABBB structure). (c) Inhomogeneous nuclear systems (AABB structure). The gyromagnetic ratio is set as $\gamma_A/\gamma_B=2$.}\label{FigS7}
\end{figure*}

The deviation density matrix of the thermal equilibrium state of a homogeneous four-qubit system
\begin{equation}\label{EqS24}
\rho_\text{eq}=\sigma_z^1+\sigma_z^2+\sigma_z^3+\sigma_z^4.
\end{equation}
is diagonalized, with the specific diagonal values shown in Figure~\ref{FigS7}(a).The state pairs that form the possible zeroth-order coherences are indicated with colored labels. It is found that the all the populations of the colored states are zero. Therefore, the $\pi/2$ line-selective pulses linking each pair of the symmetric states will not result in a nonzero value at the corresponding zeroth-order coherence positions. Consequently, we do not need an additional shaped pulse to eliminate the zeroth-order coherences.

(b). Inhomogeneous nuclear systems (ABBB structure)

The deviation density matrix of the thermal equilibrium state of the ABBB-structured four-qubit system is given by
\begin{equation}
\rho_\text{eq}=\gamma_\text{A}/\gamma_\text{B}\sigma_z^1+\sigma_z^2+\sigma_z^3+\sigma_z^4.
\end{equation}
For simplicity, we just set $\gamma_A/\gamma_B=2$. This value does not affect the following result. The corresponding diagonal elements of the density matrix are shown in Fig.~\ref{FigS7}(b). We observe that all the populations of the colored states are nonzero, indicating that the zeroth-order coherences are generated at the end of SP1. However, these zeroth-order coherence terms can be eliminated with the gradient pulse due to the different precession frequencies of the nuclei A and B in the longitudinal magnetic field.

(c). Inhomogeneous nuclear systems (AABB structure)

The deviation density matrix of the thermal equilibrium state of the AABB-structured four-qubit system is given by
\begin{equation}
\rho_\text{eq}=\gamma_\text{A}/(\gamma_\text{B}\sigma_z^1+\sigma_z^2)+\sigma_z^3+\sigma_z^4.
\end{equation}
The corresponding diagonal elements of the density matrix are shown in  Fig.~\ref{FigS7}(c).
As analyzed before, there are no nonzero values at the position $\vert0101\rangle\langle1010\vert$ and $\vert0110\rangle\langle1001\vert$. The coherence at position $\vert0011\rangle\langle1100\vert$ can be eliminated by the gradient pulse after SP1.

Thus, we conclude that our method also works for even numbers of qubits.

\section{Appendix F: Saturation Behavior of The Iteration}
\begin{figure*}
  \centering
  \includegraphics[width=1\textwidth]{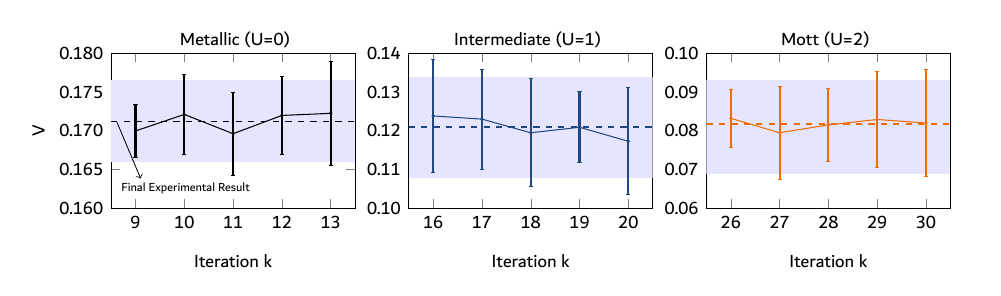}
  \caption{Last five iterative results for the bath coupling $V$ across three cases of $U$, derived from four sets of experiments. Dashed lines and shaded areas represent the averages and standard deviations of these values, respectively, indicating the final experimental values for $V$.}\label{FigS8}
\end{figure*}

As the iterations advance, the objective function $f$ and parameters 
$V$ will consistently converge towards a stable endpoint, culminating in saturation, as illustrated in Fig.~3(a-c) of the main document. When 
$U$ is held constant, the path to saturation in terms of iteration count is shaped by the initial values of the parameters and the size of each iterative adjustment. Notably, an increase in 
the value of $U$ necessitates a greater number of iterations to achieve saturation.
In practical scenarios, experimental errors may influence the saturation behavior of the iterative process. Convergence is compromised when the deviation in the cost function between successive iterations is on par with the scale of experimental errors. In our experiment, we employed the methodology outlined in [Nature 549, 242 (2017)] to determine the final experimental values of $V$.
Detailed views of the last five iterations are presented in Fig.~\ref{FigS8},  illustrating an apparent convergence influenced by experimental errors.  
We calculated the final experimental values for $V$ by averaging the last five data points, resulting in 
$0.171\pm0.005$, $0.121\pm0.013$ and $0.082\pm0.011$ for the respective $U$ cases.

\end{document}